\documentstyle[prl, aps, twocolumn]{revtex} 
\input epsf 
\epsfverbosetrue    
 
\begin{document} 
 
\title{Comment on ''Lack of Destructive Interference of Landau 
Edge States in the Quantum Hall Regime''} 
\author{J. Oswald} 
\address{Department of Physics, University of Leoben, Franz 
Josef Str. 18, A-8700 Leoben, Austria} 
\date{February 8, 1996} 
 
\maketitle 
 
In a recent Letter \cite{Mueller}, J.E. M\"uller presents a model 
calculation which demonstrates explicitly the absence of 
destructive quantum interference between edge states at the 
same sample edge in the QHE-regime. Basically the lack of 
destructive interference can be concluded from absence of 
back scattering within the same edge. Although the absence of 
back scattering is well known for a long time, a microscopic 
picture on the basis of phase coherent electron waves in 
multiply connected edge channel (EC) paths did not exist 
before. The intention of M\"ullers paper therefore was to 
present a model calculation which allows to understand how 
the quantum interference between ECs manages to avoids the 
need of back scattering. 
 
The intention of this comment is to bring up some more general 
aspects which should also be discussed in context with this 
model. The most critical point in the presented model 
calculation seems to be the choice of the initial situation.  In the 
calculation of M\"uller the initial state of the electron was 
prepared in a way that it is localized already in the incoming 
lead before the EC is split up into two phase coherent 
alternative paths. If the electron is already "on the way" to the 
dot, it is clear that there is no other way for the electron than 
being transmitted fully through the edge channel and the 
describing wavefunction has to account for this as 
demonstrated. Fig.1 shows a general situation of a quantum 
interference experiment:

\begin{figure} 
\vspace{-3mm} 
\epsfxsize=65mm                    
\epsfysize=23mm                   
\centerline{\epsfbox{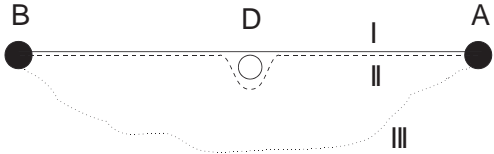}} 
\caption{Scheme of phase coherent transport including a 
barrier (D) which splits up the path into two phase coherent 
parts. The resulting two possibilities for transmission from A to 
B are represented by path I and II. Path III represents some 
alternative not particularly known paths in a real structure.} 
\label{fig1} 
\end{figure} 
 
Suppose there is an EC transmitting from (contact) A to B. It is 
split into two phase coherent parts at dot D. If the EC transport 
is phase coherent, it should be possible to describe the 
situation also by superposition of the following two phase 
coherent paths: (I) The electron leaving at A arrives at B 
without being displaced at D or (II) the electron leaving at A is 
displaced at D and arrives also at B. A realistic case should 
provide also an alternative path (labeled III) which accounts for 
other ECs or the possibility of being scattered to some probably 
present non-localized states in the bulk region of a real sample. 
In general the phase coherent transport does not allow to 
"watch" the electron without destroying the phase. All what can 
be done in an experiment is therefore to observe an electron 
leaving at A and arriving at B without knowing the exact way. 
Since the knowledge of the initial position of the electron 
implies a phase destroying event, a realistic initial location for 
the electron would be a place "outside" the phase coherent 
region at point A. But starting at A one can no longer be sure 
that the electron really enters the paths I or II because it has 
also the choice of entering III. Consequently there is only left a 
probability for the electron entering I, II or entering III. If the 
electron starts at A, there comes up another most important 
aspect which has not been considered so far: If paths I and II 
are assumed to result from an EC running from A to B, the 
transmission must include also the EC of the opposite sample 
edge. For calculating a transmission probability between A and 
B definitely the ECs at both edges are needed \cite{Buettiker}. 
Since the topic of the Letter \cite{Mueller} is clearly directed to 
the question what is going on within one particular edge, there 
is no basis for discussing any observable effects like 
transmission between the contacts or quantum interference. 
Therefore a comparison with the general aspects of a quantum 
interference experiment  like in Fig.1 must fail. This apparent 
discrepancy is probably connected to the frequently raised 
argument that considering EC-transport like a current in a real 
channel is perhaps not the whole truth of the physics behind. 
One point may be that the permanently propagating electrons 
along the edge have to be replaced by quantum mechanically 
well defined Eigenstates. An Eigenstate which is extended over 
the whole sample does not "need" interference because it has 
all information about the whole sample and therefore it "knows" 
how to arrange the associated wave function in order to 
maintain the associated (virtual) persistent edge current. On 
this background the results of M\"uller are even more important 
because he demonstrates explicitly that even in the presence 
of disorder the wavefunction of an edge electron is able to 
adjust itself in order to maintain the well defined macroscopic 
behavior of edge states. But it seems to be some how miss-
leading to characterize this effect to be a consequence of 
quantum interference.

\end{document}